# Harnessing Artificial Intelligence for Modeling Amorphous and Amorphous Porous Palladium: A Deep Neural Network Approach.

Isaías Rodríguez

Instituto de Investigaciones en Materiales, Universidad Nacional Autónoma de México Apartado Postal 70-360, Ciudad Universitaria, CDMX, México City, 04510, México.

Email: isurwars@ciencias.unam.mx

ORCID: 0000-0002-4359-2742

**ABSTRACT**

Amorphous and amorphous porous palladium are key materials for catalysis, hydrogen storage, and functional applications, but their complex structures present computational challenges. This study employs a deep neural network (DNN) trained on 33,310 atomic configurations from ab initio molecular dynamics (AIMD) simulations to model their interatomic potential. The AI-driven approach accurately predicts structural and thermal properties while significantly reducing computational costs. Validation against density functional theory (DFT) confirms its reliability in reproducing forces, energies, and structural distributions. These findings highlight AI's potential in accelerating the study of amorphous materials and advancing their applications in energy and catalysis.

**INTRODUCTION**

Bulk Metallic glasses (BMGs), often referred to as amorphous metals, have attracted significant attention in materials science and engineering due to their unique combination of mechanical, thermal, and chemical properties. Unlike their crystalline counterparts, amorphous metals do not possess a long-range ordered lattice, which leads to distinctive characteristics such as high strength, excellent corrosion resistance, and potentially superior catalytic activity in certain applications[1–4]. Among these materials, palladium-based bulk metallic glasses (Pd-BMGs) stand out because of their superior glass-forming ability, making them model systems for understanding the fundamentals of metallic glass formation and for designing advanced functional materials[5–7].

Amorphous palladium (*a*-Pd) and Pd-based alloys have a wide range of applications, including hydrogen storage, catalysis, and sensors[8–10]. For instance, the lack of grain boundaries in an amorphous structure can offer enhanced corrosion resistance and unusual catalytic sites, making amorphous Pd promising in fields such as electrochemical energy systems[11]. Moreover, palladium's well-known affinity for hydrogen absorption opens up research avenues in hydrogen separation membranes and fuel cells. Investigating the conditions under which palladium can form an amorphous phase, alongside understanding how different alloying elements affect these properties, is therefore a topic of practical and theoretical interest.

Despite the promising potential, several challenges remain in studying amorphous Pd. Computational methods, including molecular dynamics (MD) and first-principles calculations[4], have provided insights into the underlying physics. Yet, accurately simulating large amorphous systems under diverse conditions often requires considerable computational resources, and the relationship between microstructural features and macroscopic properties remains challenging to map[12].

Artificial Intelligence (AI) techniques have recently emerged as powerful tools in accelerating materials discovery and characterization[13]. By leveraging data-driven approaches, AI, and in particular machine learning (ML) can identify hidden patterns within large datasets and predict properties or phase stability with higher speed and accuracy than many traditional methods. In materials science, ML has been successfully applied to tasks such as designing new alloys, optimizing process parameters, and predicting phase diagrams[14]. In the context of metallic glasses, AI can help unveil complex correlations between composition, processing conditions, microstructure, and resultant properties relationships that may be otherwise obscured by the amorphous nature of these materials.

## METHODS

First-principles methods, such as Density Functional Theory (DFT), provide accurate energies, forces, and virials. However, they are computationally expensive when simulating large systems or long-time scales, which becomes even more critical when studying materials with defects, such as amorphicity or porosity. Previous work has demonstrated that systems comprising a few hundred atoms are sufficient to accurately capture the structural and electronic properties of amorphous materials[15–17]. However, thousands of atoms are frequently required to represent porous materials reliably[18]. Although classical interatomic potentials such as Lennard-Jones (L-J), the embedded atom method (EAM), or Tersoff can handle systems with thousands of atoms, they frequently lack the accuracy needed to predict the properties of complex materials.

Machine learning (ML) potentials help bridge this gap by learning to predict energies and forces with accuracy comparable to first-principles methods, but at a computational cost closer to that of classical force fields. In particular, deep learning makes it possible for a trained neural network to reach the same level of accuracy as the DFT calculations used for its training, thus allowing it to replace traditional force fields in classical molecular dynamics (CMD) simulations. Achieving high performance with an ML model, however, typically requires a large training dataset, highlighting the importance of acquiring an extensive set of DFT calculations.

In this work, we performed *ab initio* molecular dynamics (AIMD) simulations of both amorphous and amorphous porous palladium to generate a dataset of 33,310 atomic configurations[4,12]. All calculations were carried out using the CASTEP software[19], which is part of the BIOVIA Materials Studio suite[20]. For these AIMD simulations, we employed a plane-wave energy cutoff of 330 eV, within the generalized gradient approximation (GGA). The amorphous palladium[4] structure were calculated with the Perdew-Burke-Ernzerhof (PBE)[21] parametrization for the exchange-correlation functional, While the amorphous porous palladium samples[12] were calculated with the PBE for solids parametrization (PBEsol)[22]. Ultrasoft pseudopotentials were used for the core treatment. This methodology has been shown to accurately capture the structural, electronic, magnetic and thermal properties of both amorphous and amorphous porous palladium.

Therefore, the deep learning neural network developed in this work can accurately capture the structural properties of amorphous and amorphous porous palladium up to the mesoscale, surpassing the size limitations typically encountered with standard AIMD simulations.

The deep neural network (DNN) model used to describe the interatomic potential consists of three hidden layers, each containing 240 neurons and using the tanh activation function. The model was trained using the DeePMD-kit package [23] on 33,310 randomly selected *ab-initio* configurations, each consisting of 216-atom supercells generated via the *undermelt-quench* approach[15]. These AIMD simulations involve a heating phase of 100 steps, where the temperature is raised from 300 K to 1500 K, followed by a cooling phase from 1500 K down to ~0 K. A subsequent geometry optimization (GO), requiring around 1000 steps on average, is then performed. This procedure was applied to systems with porosities of 0%, 25%, 50%, 67.5%, 75%, and 82.5%.

Subsequently, the trained DNN was employed to generate large amorphous and amorphous porous palladium systems containing 85,184 atoms using the LAMMPS package [24], following a protocol similar to the original *undermelt-quench* method. Structural properties were then calculated using the Correlation software, previously developed by the Rodríguez *et. al.*[25].

## RESULTS AND DISCUSSION

In this study, two robust validation procedures were implemented to evaluate the performance of the DNN model. The first procedure involved a direct comparison of the forces and energies predicted by the DNN model against those from the DFT dataset. A total of 6,660 randomly selected samples were analyzed, with the results presented in **Figure 1**, showcasing the accuracy and reliability of the model.

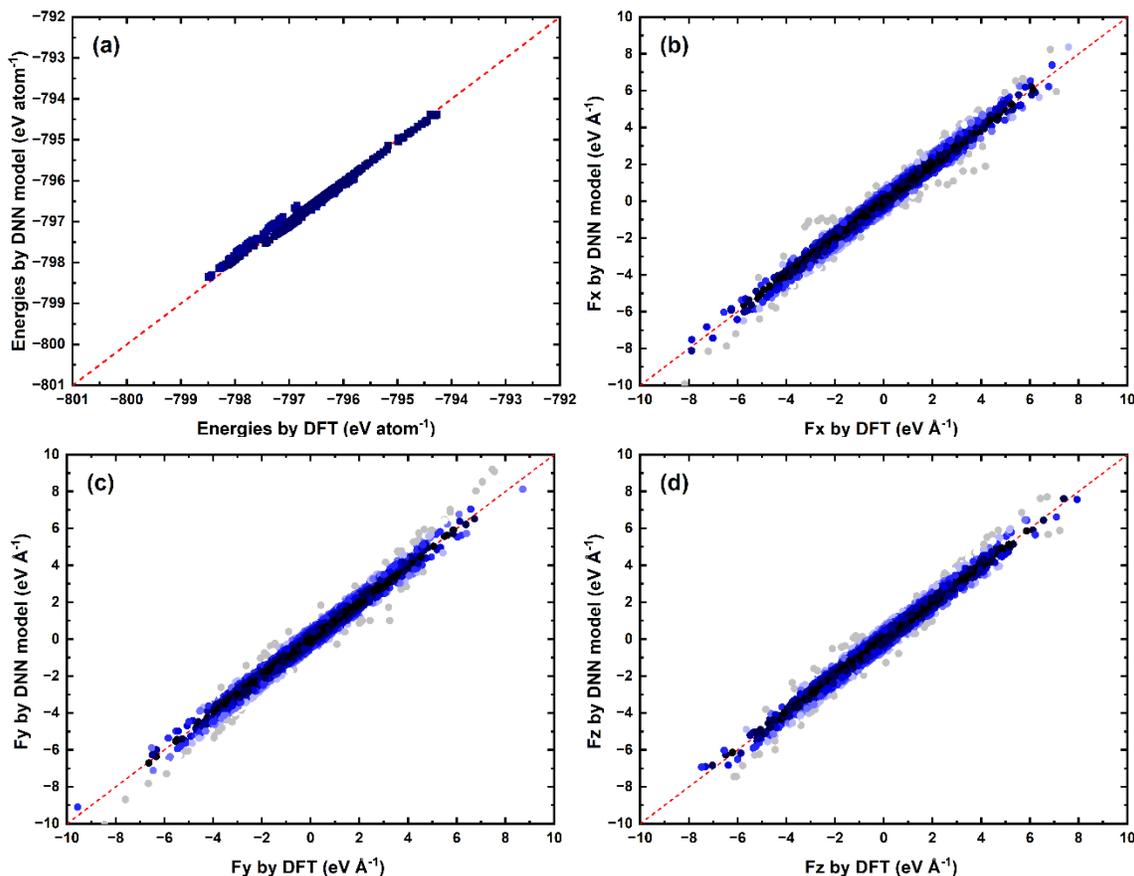

**Figure 1.** Validation of the DNN model against the DFT dataset: (a) Energies per atom; (b), (c), and (d) Forces per atom in the x, y, and z directions, respectively. The identity line, in red, represents an almost perfect agreement between predictions and reference values.

**Figure 1** illustrates that the majority of data points align closely with the identity line, signifying a high level of agreement between the predictions of the DNN model and the DFT results. This strong correlation underscores the reliability of the DNN potential in accurately predicting forces and energies for both amorphous and amorphous porous palladium. Notably, the maximum force discrepancies, ranging from above 6 eV up to 10 eV/Å, were observed near the upper range of absolute values, primarily in structures undergoing significant internal strain during transitions from crystalline to amorphous phases.

Similarly, the largest energy discrepancies were observed in the range between 799 eV and -797 eV, indicating a higher order of deviation in these energy values. These discrepancies are strongly correlated with crystalline structures, suggesting that the DNN model captures the transition states with high accuracy but exhibits limitations in fully reproducing the energy landscape of highly ordered systems. This may indicate that more crystalline structure samples need to be incorporated into the model to fine-tune the DNN potential, improving its ability to accurately represent crystalline structures and their associated energy distributions.

The second validation process involved performing detailed simulations of amorphous and amorphous porous palladium systems with porosities of 50%, and 75%. These simulations aimed to assess the performance of the DNN potential by comparing its predictions for the pair correlation function (PDF) and plane-angle distribution (PAD) against benchmark values reported in the literature[12]. Figures 2 and 3 showcase the results, underscoring the model's precision in reproducing the structural characteristics of amorphous palladium.

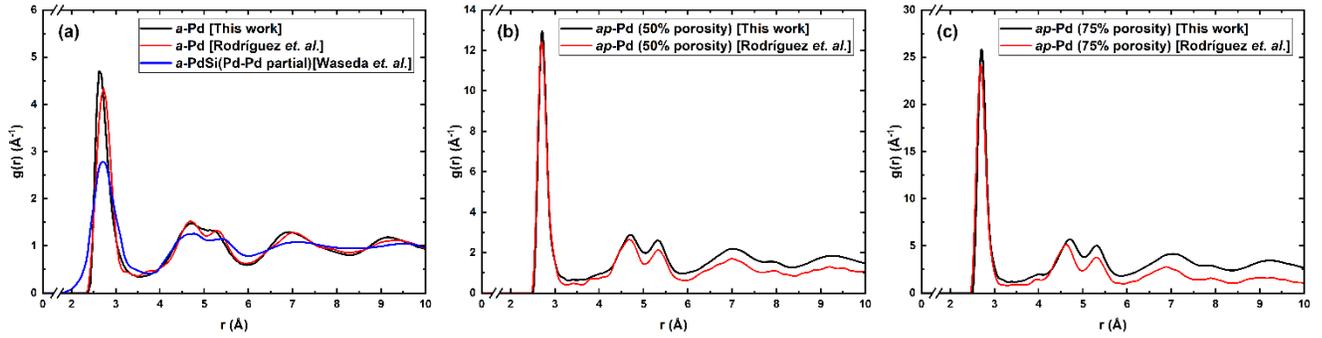

**Figure 2.** Pair distribution function (PDF) comparison for the simulated amorphous and amorphous porous palladium systems. The PDFs from this work (black) are shown alongside those reported by Rodríguez et al. [4,12] (red) for three porosity levels: (a) 0%, (b) 50%, and (c) 75%. Additionally, the experimental results from Waseda et al.[26] (blue) are included for the non-porous (0% porosity) case.

**Figure 2** highlights the remarkable agreement in the first coordination sphere across all cases, including the comparison with the experimental results from Waseda et al. in **Figure 2(a)**[26]. For the second neighbors shell, the alignment of peak positions is notable in all samples, and the metallic bimodality of the second neighbors is consistently observed. However, a decrease below 1 in the PDF for the DFT calculations is apparent in **Figure 2(b) and 2(c)**, a phenomenon previously reported for porous materials and likely associated with the pore size effects of the DFT dataset[12,27].

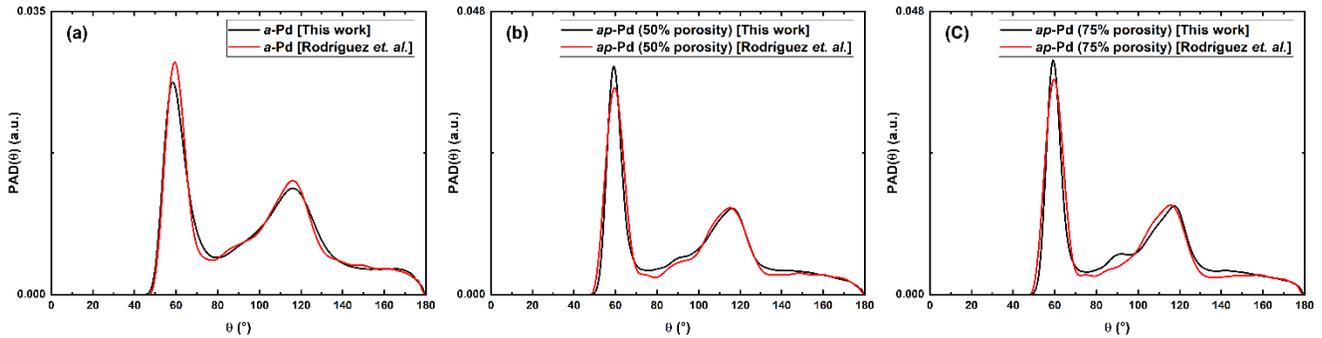

**Figure 3.** Plane-Angle Distribution (PAD) comparison for simulated amorphous and amorphous porous palladium systems. The PADs from this study (black) are compared to those reported by Rodríguez et al.[4,12] (red) for three porosity levels: (a) 0%, (b) 50%, and (c) 75%.

**Figure 3** demonstrates that the Plane-Angle Distribution (PAD) is remarkably consistent across all samples, as expected, given that PADs are predominantly determined by the first coordination shells. This observation aligns with **Figure 2**, where the first peak of the PDFs shows an almost perfect match with the DFT dataset. These results demonstrate the DNN model's capability to accurately capture structural variations across different porosities.

To further highlight the impact of porosity on the structural characteristics of amorphous materials, a series of 8,000-atom samples were simulated using the same *undermelt-quench* process as the 85,184-atom samples employed for DNN model validation. These simulations focused on amorphous porous palladium (*ap*-Pd) with porosities of 50% and 75%. The resulting structures, depicted in Figure 4, provide comprehensive insights into the distribution, connectivity, and morphological variations of pores at different porosity levels. Additionally, these simulations underline the DNN model's precision in capturing local atomic arrangements and reproducing intricate structural features consistently across various porosities.

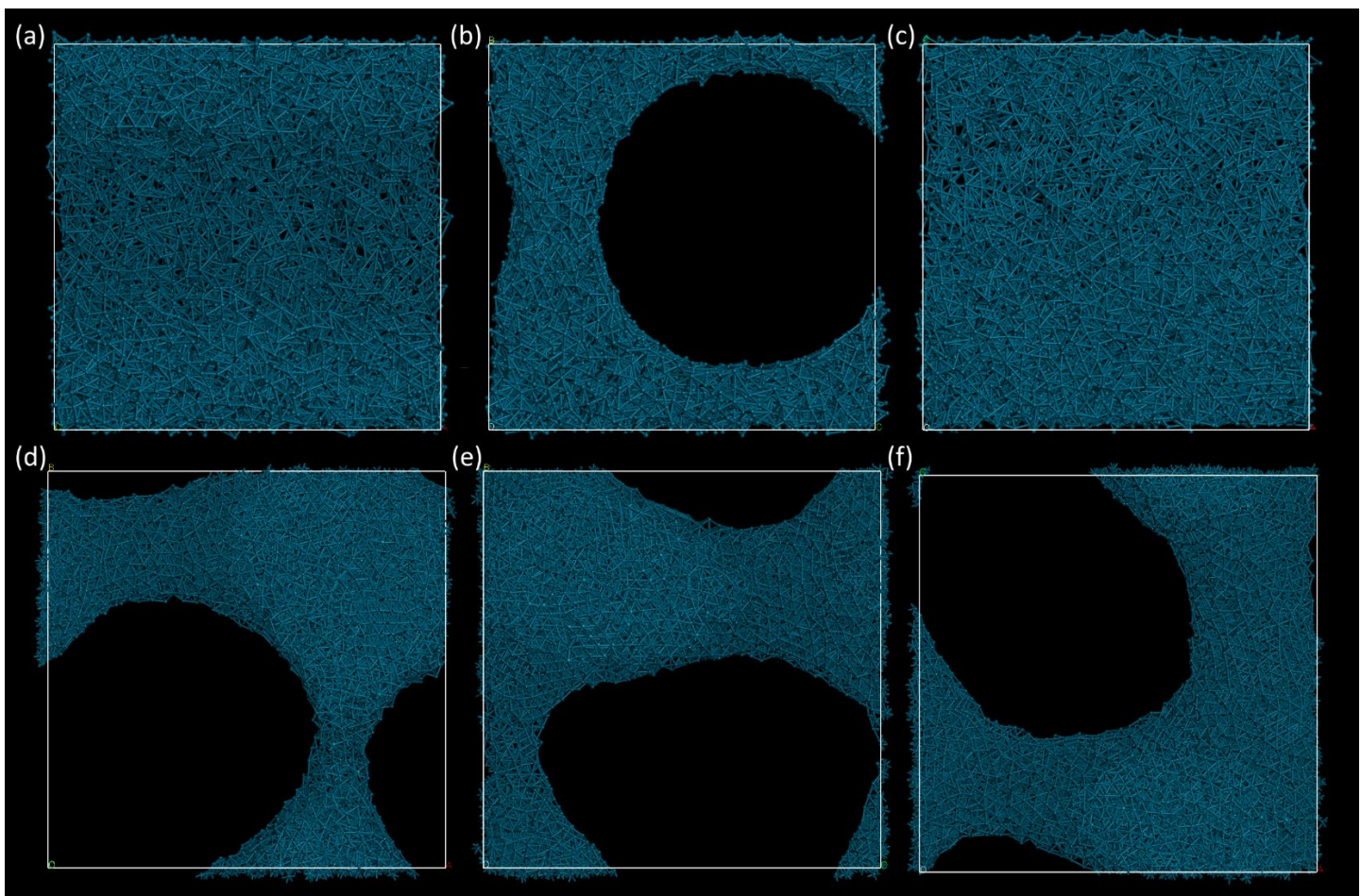

**Figure 4**. Visualization of amorphous porous palladium structures from an 8,000-atom simulation using the DNN model. The projections highlight the material's microstructural features at two porosity levels: 50% porosity with XY (a), YZ (b), and XZ (c) projections, and 75% porosity with XY (d), YZ (e), and XZ (f) projections. These images show the pore distribution and connectivity, as well as local atomic arrangements.

There appear to be two distinct types of porous structures in palladium. The first type consists of nearly cylindrical through-pores, a characteristic extensively documented in the literature through experimental studies on sensors and catalytic materials[10,28], as well as theoretical simulations[12]. These through-pores typically exhibit diameters ranging from approximately 30 nm to several hundred nanometers, depending on the fabrication method and material processing conditions. In **Figure 4(b)**, a cylindrical pore with a diameter of approximately 70 nm is prominently visible along the X-axis, while no noticeable porosity is observed along the Y and Z axes, as depicted in **Figures 4 (a) and (c)**. This anisotropic distribution of porosity suggests a directional formation mechanism, which could be influenced by external processing conditions such as applied stress, temperature gradients, or templating techniques. Such insights further validate the DNN model for amorphous porous palladium systems.

The second type of structure forms a mesh of interconnected backbone material, with through-pores visible in multiple directions, resembling a foam-like architecture, as can be seen in **Figure 4(d), (e) and (f)**. This low-density ap-Pd foam has been extensively studied in experimental research due to its high surface-to-mass ratio, making it particularly advantageous for applications in catalysis and hydrogen storage[29,30]. The foam structure facilitates enhanced catalytic activity by increasing the accessibility of active sites, while its porous network supports efficient hydrogen absorption and release. Such unique structural characteristics make these materials promising candidates for next-generation energy storage and conversion technologies.

# CONCLUSIONS

In summary, this work successfully developed a deep neural network potential model capable of accurately reproducing the structural properties of both amorphous and amorphous porous palladium. The effectiveness of the model highlights the potential of deep neural networks to reliably simulate complex structures in amorphous materials, including through-pores and low-density foams, also known as metallic aerogels. By achieving high fidelity in replicating experimental and theoretical benchmarks, this approach demonstrates the viability of AI-driven methods for studying a wide range of disordered materials. Furthermore, the ability to simulate pore connectivity, distribution, and local atomic arrangements across varying porosity levels underscores its applicability in the design and optimization of materials for catalysis, hydrogen storage, and energy conversion technologies. This work paves the way for future studies leveraging AI to explore other complex materials and phenomena, offering a powerful tool for advancing materials science.

**Acknowledgements**

The author wants to acknowledge Professor Ariel A. Valladares for invaluable mentorship and support. Special thanks to Renela M. Valladares, Alexander Valladares and David Hinojosa-Romero for insightful discussions and continuous encouragement. María Teresa Vázquez and Oralia Jiménez provided essential administrative assistance. Technical support for computing resources was generously provided by Alberto López† and Alejandro Pompa at IIM-UNAM.

**Author Contributions**

The sole author conceived and designed the study, collected the dataset, coded the deep neural network, performed the analyses, and wrote the manuscript.

**Funding**

The author expresses his gratitude to Consejo Nacional de Humanidades, Ciencia y Tecnología (CONAHCyT) for the postdoctoral fellowship. This project was funded by Dirección General de Personal Académico de la Universidad Nacional Autónoma de México (DGAPA-UNAM, PAPIIT) under Grant No. IN118223. Simulations were partially conducted at the Supercomputing Center at Dirección General de Cómputo y de Tecnologías de Información y Comunicación (DGTIC-UNAM) through project LANCAD-UNAM-DGTIC-131.

**Data Availability Statement**

The data supporting the findings of this study are available from the corresponding author upon reasonable request. The developed code, trained neural network, and associated datasets will be publicly released in a forthcoming publication to ensure transparency and reproducibility of the results.

**Conflict of Interest**
The author declares no conflict of interest.